\documentclass[12pt]{article}
\usepackage{times,amsfonts,amssymb,amsmath,bbm,stmaryrd,amscd,latexsym}
\usepackage[dvips]{graphicx}
\def\de{\delta}

\def\k{\kappa}

\def\om{\omega}

\def\Om{\Omega}

\def\S{\Sigma}

\def\sfrac#1#2{{\textstyle\frac{#1}{#2}}}

\def\ph{\phantom{-}}
\def\ic{\mathrm{i}}
\def\eu{\mathrm{e}}
\def\diff{\mathrm{d}}
\def\tr{\mathrm{tr}}
\def\pa{\partial}
\def\we{\wedge}
\def\>{\rangle}
\def\<{\langle}
\def\+{\dagger}
\def\={\ =\ }
\newcommand{\C}{\mathbb C}
\newcommand{\R}{\mathbb R}

\newcommand{\fk}{\mathfrak{k}}
\newcommand{\fh}{\mathfrak{h}}
\newcommand{\fm}{\mathfrak{m}}
\newcommand{\unity}{{\mathbbm{1}}}
\newcommand{\J}{{J}}
\newcommand{\hP}{\hat{\Phi}}
\newcommand{\be}{\begin{equation}}
\newcommand{\ee}{\end{equation}}
\newcommand{\bea}{\begin{eqnarray}}
\newcommand{\eea}{\end{eqnarray}}
\newcommand{\bal}{\begin{aligned}}
\newcommand{\eal}{\end{aligned}}

\newcommand{\und}{\qquad{\text{and}}\qquad}
\begin{document}
\begin{titlepage}
\begin{flushright}      
ITP--UH--03/12\\        
\end{flushright}        

\vskip 2.0cm

\begin{center}
{\Large\bf Instantons and Chern-Simons Flows in 6, 7 and 8 dimensions}

\vskip 1.5cm

{\Large \ Olaf Lechtenfeld}

\vskip 0.5cm

{\it Institut f\"ur Theoretische Physik, Leibniz Universit\"at Hannover}\\
{\it Appelstra\ss{}e 2, 30167 Hannover, Germany}\\
{URL: www.itp.uni-hannover.de/\~{}lechtenf}

\vskip 0.2cm
and
\vskip 0.2cm

{\it Centre for Quantum Engineering and Space-Time Research}\\
{\it Leibniz Universit\"at Hannover}\\
{\it Welfengarten 1, 30167 Hannover, Germany}

\end{center}
\vskip 1.5cm

\begin{abstract} \noindent
The existence of $K$-instantons on a cylinder $M^7=\R_\tau\times\sfrac{K}{H}$ 
over a homogeneous nearly K\"ahler 6-manifold $\sfrac{K}{H}$ requires 
a conformally parallel or a cocalibrated $G_2$-structure on~$M^7$. 
The generalized anti-self-duality on~$M^7$ implies a Chern-Simons flow
on~$\sfrac{K}{H}$ which runs between instantons on the coset.
For $K$-equivariant connections, the torsionful Yang-Mills equation reduces 
to a particular quartic dynamics for a Newtonian particle on~$\C$.
When the torsion corresponds to one of the $G_2$-structures, this dynamics
follows from a gradient or hamiltonian flow equation, respectively.
We present the analytic (kink-type) solutions and plot numerical non-BPS 
solutions for general torsion values interpolating between the instantonic ones.
\end{abstract}

\vfill

Talk presented at SQS-11 during 18--23 July, 2011, at JINR, Dubna, Russia

\end{titlepage}


\newpage

\section{Introduction}

Yang-Mills instantons exist dimensions~$d$ larger than four only when there is
additional geometric structure on the manifold~$M^d$ (besides the Riemannian one).
In order to formulate generalized first-order anti-self-duality conditions which 
imply the second-order Yang-Mills equations (possibly with torsion), $M^d$ must be 
equipped with a so called $G$-structure, which is a globally defined but not
necessarily closed $(d{-}4)$-form~$\Sigma$, so that the weak holonomy group
of~$M^d$ gets reduced.

Instanton solutions in higher dimensions are rare in the literature. 
In the mid-eighties, Fairlie and Nuyts and also Fubini and Nicolai discovered the
Spin(7)-instanton on~$\R^8$. Eight years later, a similar $G_2$-instanton on~$\R^7$
was found by Ivanova and Popov and also by G\"{u}naydin and Nicolai. Our recent work
shows that these so called octonionic instantons are not isolated but embedded into 
a whole family living on a class of conical non-compact manifolds~\cite{Gemmer2011}.

The string vacua in heterotic flux compactifications contain non-abelian gauge fields
which in the supergravity limit are subject to Yang-Mills equations with torsion~$\cal H$ 
determined by the three-form flux.
Prominent cases admitting instantons are AdS$_{10-d}\times M^d$, where $M^d$ is
equipped with a $G$-structure, with $G$ being SU(3), $G_2$ or Spin(7) for $d=6$, $7$ 
or $8$, respectively. Homogeneous nearly K\"ahler 6-manifolds $\sfrac{K}{H}$ and
(iterated) cylinders and (sine-)cones over them provide simple examples, for which all
$K$-equivariant Yang-Mills connections can be constructed~\cite{Harland2009, Bauer2010}. 
Natural choices for the gauge group are $K$ or~$G$.

Clearly, the Yang-Mills instantons discussed here serve to construct heterotic string
solitons, as was first done in 1990 by Strominger for the gauge five-brane.
It is therefore of interest to extend our new instantons to solutions of (string-corrected)
heterotic supergravity and obtain novel string/brane vacua 
\cite{Lechtenfeld2010, Chatzistavrakidis2012, Gemmer2012}.

In this talk, I present the construction for the simplest case of a cylinder over a
compact homogeneous nearly K\"ahler coset~$\sfrac{K}{H}$, which allows for a conformally
parallel or a cocalibrated $G_2$-structure. I display a family of non-BPS Yang-Mills 
connections, which contain two instantons at distinguished parameter values corresponding
to those $G_2$-structures. In these two cases, anti-self-duality implies a Chern-Simons
flow on~$\sfrac{K}{H}$.

Finally, I must apologize for the omission -- due to page limitation -- of all relevant 
literature besides my own papers on which this talk is based. The reader can find all 
references therein.

\section{Self-duality in higher dimensions}

The familiar four-dimensional anti-self-duality condition 
for Yang-Mills fields~$F$
may be generalized to suitable $d$-dimensional Riemannian manifolds $M$,
\be \label {ASD}
*F \= -\Sigma\wedge F \qquad\text{for}\qquad
F=\diff A + A\wedge A \qquad\text{and}\qquad
\Sigma\in\Lambda^{d-4}(M)\ ,
\ee
if there exists a geometrically natural $(d{-}4)$-form~$\Sigma$ on~$M$.
Applying the gauge-covariant derivative \ $D=\diff+[A,\cdot]$ \ it follows that
\be \label{YMT}
D{*}F\ +\ \diff\Sigma\wedge F \= 0 \qquad\Longleftrightarrow\qquad
\text{Yang-Mills with torsion} \quad
{\cal H}=*\diff\Sigma\in\Lambda^3(M)\ .
\ee
This torsionful Yang-Mills equation extremizes the action
\be \label{fullaction}
S_{\text{YM}}+S_{\text{CS}}\=
\int_M\!\tr\bigl\{ F\wedge*F\ +\ (-)^{d-3}\Sigma\wedge F\wedge F\bigr\}\=
\int_M\!\tr\bigl\{ F\wedge*F\ +\ 
\sfrac12\diff\Sigma\wedge\bigl(A\,\diff A+\sfrac23 A^3\bigr)\bigr\}\ .
\ee
Related to this generalized anti-self-duality is the
gradient Chern-Simons flow on $M$,
\be
\frac{\diff A}{\diff\tau}\=\frac{\delta}{\delta A} S_{\text{CS}}
\= *\bigl(\diff\Sigma\wedge F\bigr)\ \sim\ *\diff\Sigma\,\lrcorner\,F\ .
\ee
In fact, this equation follows from generalized 
anti-self-duality on the cylinder $\widetilde{M}=\R_\tau{\times}M$
over $M$ (in the $A_\tau{=}0$ gauge).

The question is therefore: 
Which manifolds admit a global $(d{-}4)$-form? 
And the answer is:
$G$-structure manifolds, i.e.~manifolds with a weak special holonomy.
Let me give the key cases we shall encounter in this talk:
\begin{center}
\begin{tabular}{|lccccl|}
\hline
$d$ & $G$     & $\Sigma$ & cases & example & structure \\ \hline 
$6$ & SU(3)   & $\omega$ & K\"ahler $\vphantom{\Big|}$ &
$\C P^3$ &
$\diff\omega=0$ \\[2pt]
$6$ & SU(3)   & $\omega$ & nearly K\"ahler &
$S^6=\frac{G_2}{\text{SU}(3)}$ &
$\diff\omega\sim\text{Im}\Omega$\,,\ 
$\diff\text{Re}\Omega\sim\omega^2$ \\[8pt]
$7$ & $G_2$   & $\psi$   & conf.\ parallel $G_2$ & 
$\R_\tau{\times}\,\text{nearly K\"ahler}$ & 
$\diff\psi\sim\psi{\wedge}\diff\tau$\,,\ 
$\diff{*}\psi\sim-{*}\psi{\wedge}\diff\tau$ \\[2pt]
$7$ & $G_2$   & $\psi$   & nearly parallel $G_2$ & 
$X_{k,\ell}=\frac{\text{SU}(3)}{\text{U}(1)_{k,\ell}}$ & 
$\diff\psi\sim*\psi\ \Rightarrow\ \diff{*}\psi=0$ \\
$7$ & $G_2$   & $\psi$   & parallel $G_2$ &
$\text{cone}(\text{nearly K\"ahler})$ & $\diff\psi=0=\diff{*}\psi$ \\[8pt]
$8$ & Spin(7) & $\Sigma$ & parallel Spin(7) & 
$\R_\tau{\times}\,\text{parallel}\,G_2$ & 
$\diff\Sigma=0$\,,\ $*\Sigma=\Sigma$ \\
\hline
\end{tabular}
\end{center}
Some of those cases are related via the following scheme,
with examples in square brackets.\\[8pt]
\begin{minipage}{12cm}
\includegraphics[width=12cm]{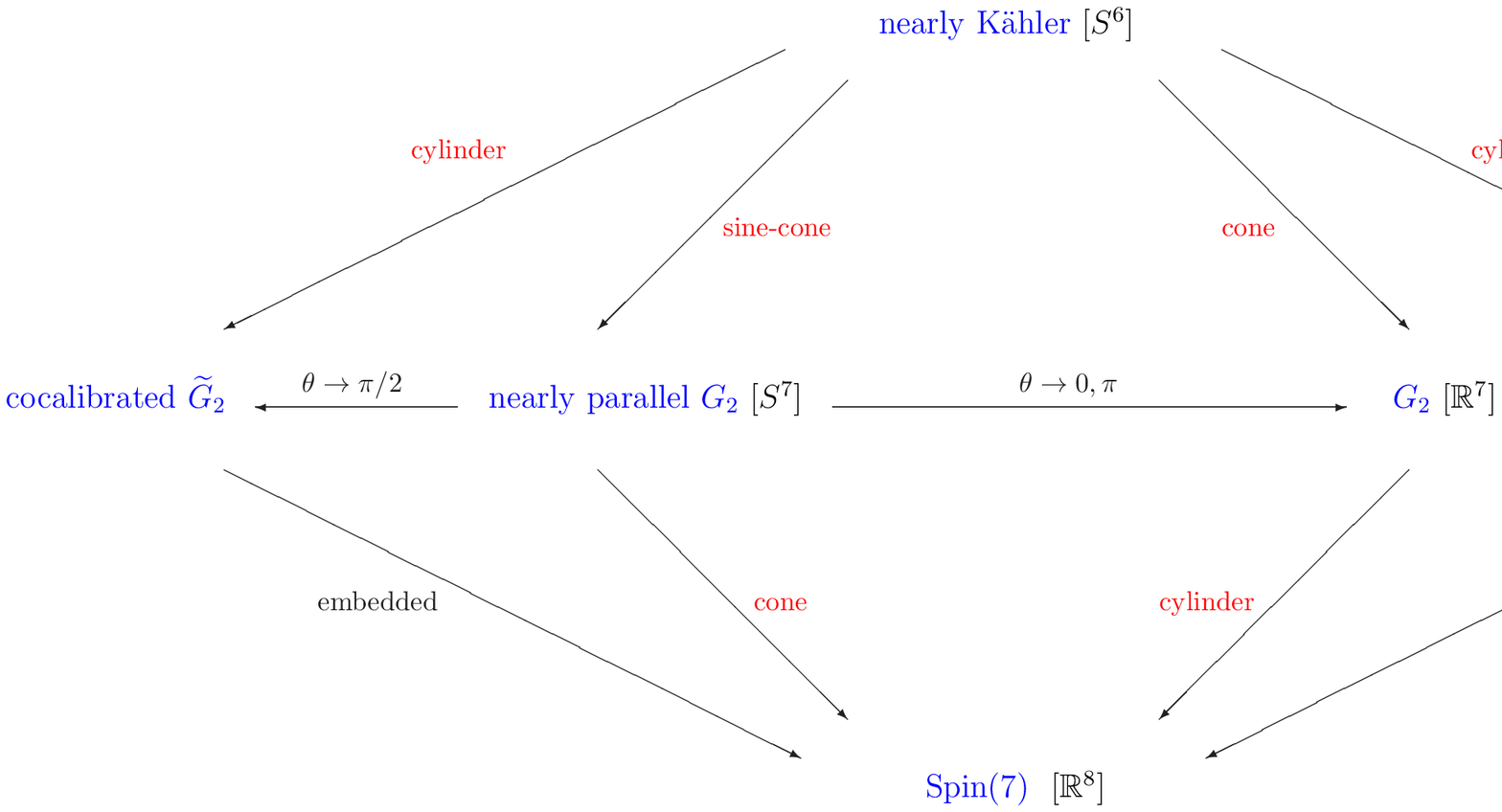}
\end{minipage}\\[16pt]
For this talk I shall consider (reductive non-symmetric) coset spaces 
$M{=}\frac{K}{H}$ in $d{=}6$ as well as cylinders and cones over them.
In all these cases, the gauge group is chosen to be $K$.

\section{Six dimensions: nearly K\"{a}hler coset spaces}

All known compact nearly K\"{a}hler 6-manifolds~$M^6$ are nonsymmetric 
coset spaces $K/H$: 
\be
S^6=\sfrac{G_2}{\text{SU}(3)}\ ,\quad
\sfrac{\text{Sp}(2)}{\text{Sp}(1){\times}\text{U}(1)}\ ,\quad
\sfrac{\text{SU}(3)}{\text{U}(1){\times}\text{U}(1)}\ ,\quad
S^3{\times}S^3=
\sfrac{\text{SU}(2){\times}\text{SU}(2){\times}\text{SU}(2)}{\text{SU}(2)}\ .
\ee
The coset structure $H\lhd K$ implies the decomposition
\be
\text{Lie}(K)\equiv\fk=\fh\oplus\fm \quad\text{with}\quad 
\fh\equiv\text{Lie}(H)\quad\text{and}\quad [\fh,\fm]\subset\fm\ .
\ee
Interestingly, the reflection automorphism of symmetric spaces gets generalized
to a so called tri-symmetry automorphism $S:K\to K$ with $S^3=\text{id}$ implying
\be
s:\fk\to\fk \qquad\text{with}\qquad 
s|_\fh=\unity \quad\text{and}\quad 
s|_\fm=-\sfrac12+\sfrac{\sqrt{3}}{2}\,\J =\exp\{\sfrac{2\pi}3\J\}\ ,
\ee
effecting a $\sfrac{2\pi}{3}$ rotation on~$TM^6$.
I pick a Lie-algebra basis
\be
\{I_{a=1,\ldots,6}\,,\,I_{i=7,\ldots,\smash{\text{dim}G}}\}
\qquad\text{with}\qquad [I_a,I_b]=f^i_{ab}I_i+f^c_{ab}I_c\ ,
\ee
involving the structure constants $f^{\bullet}_{ab}$.
The Cartan-Killing form then reads
\be
\<\cdot,\cdot\>_\fk \= -\tr_\fk(\text{ad}(\cdot)\circ\text{ad}(\cdot))
\= 3\,\<\cdot,\cdot\>_\fh \= 3\,\<\cdot,\cdot\>_\fm \=\mathbbm{1}\ .
\ee
Expanding all structures in a basis of canonical one-forms $e^a$ framing $T^*(G/H)$, 
\be
g\=\de_{ab}\,e^ae^b\ ,\qquad 
\om\=\sfrac12\J_{ab}\,e^a\we e^b\ ,\qquad 
\Om\=-\sfrac{1}{\sqrt{3}}(f+\ic\J f)_{abc}\,e^a\we e^b\we e^c\ ,
\ee
we see that the almost complex structure $(J_{ab})$ and the structure constants
$f_{abc}$ rule everything.

Nearly K\"ahler 6-manifolds are special in that the torsion term in~(\ref{YMT}) 
{\em vanishes by itself\/}! What is more, this property is actually {\em equivalent\/} 
to the generalized anti-self-duality condition~(\ref{ASD}):
\be \label{NKacc}
*F \= -\omega\wedge F \qquad\Longleftrightarrow\qquad
0\=\diff\omega\wedge F\ \sim\ \text{Im}\Omega\wedge F
\qquad\Longleftrightarrow\qquad\text{DUY equations}\ ,
\ee
where the Donaldson-Uhlenbeck-Yau (DUY) equations~\footnote{
also known as `hermitian Yang-Mills equations'}
state that 
\be
F^{2,0}=F^{0,2}=0 \qquad\text{and}\qquad \omega\lrcorner F=0\ .
\ee
Another interpretation of this anti-self-duality condition is that is projects~$F$
to the 8-dimensional eigenspace of the endomorphism $*(\omega\wedge\cdot)$ with
eigenvalue~$-1$, which contains the part of $F^{1,1}$ orthogonal to~$\omega$.
The equations~(\ref{NKacc}) imply also $\text{Re}\Omega\wedge F=0$ and
the (torsion-free) Yang-Mills equations $D{*}F=0$. 
Clearly, they seperately extremize both $S_{\text{YM}}$ and $S_{\text{CS}}$
in~(\ref{fullaction}), but of course yield only BPS-type classical solutions.
In components the above relations take the form
\bea
\sfrac12\epsilon_{abcdef}F_{ef}\=-\J_{[ab}F_{cd]}
\qquad\Longleftrightarrow\qquad
0\=f_{abc}F_{bc}\ ,  \\[2pt]
\Longrightarrow \qquad
\omega_{ab}F_{ab}=0\ ,\quad (\J f)_{abc}F_{bc}=0\ ,\quad D_a F_{ab}=0\ .
\eea
I notice that each Chern-Simons flow \ $\dot A_a\sim f_{abc}F_{bc}$ \ on $M^6$ 
ends in an instanton.

Let me look for $K$-equivariant connections~$A$ on~$M^6$.
If I restrict their value to~$\fh$, the answer is unique: the only
`$H$-instanton' is the so called canonical connection
\be
A^{\text{can}}=e^i\,I_i \qquad\longrightarrow\qquad
F^{\text{can}}=-\sfrac12 f^i_{ab}\,e^a{\we}e^b\,I_i\ ,
\ee
where $e^i=e^i_a e^a$.
Generalizing to `$K$-instantons', I extend to
\be \label{ansatz6}
A=e^i\,I_i+e^a\,\Phi_{ab}\,I_b
\qquad\text{with ansatz}\qquad
\bigl(\Phi_{ab}\bigr)=:\Phi=\phi_1\unity+\phi_2\J\ ,
\ee
which is in fact general for $G_2$ invariance on $S^6$.
Its curvature is readily computed to
\be
F_{ab}\= F_{ab}^{1,1}+F_{ab}^{2,0\oplus0,2}\=
(|\Phi|^2{-}1)\,f^i_{ab}\,I_i\ +\ [(\bar\Phi^2{-}\Phi)\,f]_{abc}\,I_c
\ee
and displays the tri-symmetry invariance under 
$\Phi\to\exp\{-\sfrac{2\pi}3\J\}\Phi$.
The solutions to the BPS conditions~(\ref{NKacc}) are
\be
\bar\Phi^2=\Phi\qquad\Longrightarrow\qquad
\Phi=0\qquad\text{or}\qquad\Phi=\exp\{\sfrac{2\pi k}3\J\}
\quad\text{for}\quad k=0,1,2\ ,
\ee
which yields three flat $K$-instanton connections besides
the canonical curved one,
\be
A^{(k)}=e^i\,I_i+e^a\,(s^k I)_a \qquad\text{and}\qquad 
A^{\text{can}}=e^i\,I_i\ .
\ee

\section{Seven dimensions: cylinder over nearly K\"{a}hler cosets}

Let me step up one dimension and consider 7-manifolds~$M^7$ with weak $G_2$ holonomy
associated with a $G_2$-structure three-form~$\psi$.
Here, the 7 generalized anti-self-duality equations project $F$ onto the $-1$~eigenspace
of $*(\psi\wedge\cdot)$, which is 14-dimensional and isomorphic to the Lie algebra of~$G_2$,
\be \label{G2ASD}
*F \= -\psi\wedge F \qquad\Longleftrightarrow\qquad
*\psi\wedge F \= 0\qquad\Longleftrightarrow\qquad
\psi\lrcorner F\= 0\ ,
\ee
providing an alternative form of the condition. In components, it reads
\be
\sfrac12\epsilon_{abcdefg}F_{fg}\=-\psi_{[abc}F_{de]} \qquad\Longleftrightarrow\qquad
0\=\psi_{abc}F_{bc} \ .
\ee
For the parallel and nearly parallel $G_2$ cases, the previous accident~(\ref{NKacc}) recurs,
\be \label{NPG2acc}
\diff\psi\ \sim\ *\psi \qquad\Longrightarrow\qquad
\diff\psi\wedge F\= 0 \qquad\Longrightarrow\qquad D*F\= 0\ ,
\ee
and the torsion decouples. 
Note that on a general weak $G_2$-manifold there are two different flows,
\be
\frac{\diff A(\sigma)}{\diff\sigma} \= *\diff\psi\,\lrcorner\,F(\sigma) \und 
\frac{\diff A(\sigma)}{\diff\sigma} \= \psi\,\lrcorner\,F(\sigma)
\qquad\text{for}\quad\sigma\in\R\ ,
\ee
which coincide in the nearly parallel case. The second flow ends in an instanton on~$M^7$.

In this talk I focus on cylinders \
$M^7=\R_\tau\times\sfrac{K}{H}$
over nearly K\"ahler cosets, with a metric \
$g=(\diff\tau)^2+\de_{ab}\,e^ae^b$, on which I study the Yang-Mills equation
with a torsion given by
\be \label{ktorsion}
*{\cal H}=\sfrac13 \k\,\diff\omega\wedge\diff\tau
\qquad\Longleftrightarrow\qquad T_{abc}=\k\,f_{abc}
\ee
with a real parameter~$\k$.
We shall see that for special values of~$\k$ my torsionful Yang-Mills equation
\be \label{YMT7}
D{*}F+\sfrac13\k\,\diff\omega\wedge\diff\tau\wedge{F}\=0
\ee
descends from an anti-self-duality condition~(\ref{G2ASD}).

Taking the $A_0{=}0$ gauge and
borrowing the ansatz~(\ref{ansatz6}) from the nearly K\"ahler base, I write
\be \label{ansatz7}
A_a=e^i_a I_i+[\Phi(\tau)\,I]_a \qquad\Rightarrow\qquad
F_{0a}=[\dot\Phi\,I]_a \ ,\quad F_{ab}=
(|\Phi|^2{-}1)\,f^i_{ab}\,I_i\ +\ [(\bar\Phi^2{-}\Phi)\,f]_{abc}\,I_c
\ee
which depends on a complex function~$\Phi(\tau)$ (values in the $(\unity,J)$ plane).
Sticking this into~(\ref{YMT7}) and computing for a while, one arrives at
\be \label{phieom}
\ddot\Phi\=(\k{-}1)\Phi\ -\ (\k{+}3)\bar\Phi^2\ +\ 4\bar\Phi\Phi^2
\ =:\ \sfrac13 \sfrac{\pa V}{\pa\bar\Phi}\ .
\ee
Nice enough, I have obtained a $\phi^4$ model with an action
\be
S[\Phi]\ \sim\int_{\R}\!\diff\tau\,\bigl\{ 3|\dot\Phi|^2 + V(\Phi) \bigr\}
\quad\text{for}\quad
V(\Phi)\= (3{-}\k)\ +\ 3(\k{-}1)|\Phi|^2\ -\ (3{+}\k)(\Phi^3{+}\bar\Phi^3)\ +\ 6|\Phi|^4
\ee
devoid of rotational symmetry (for $\k{\neq}{-}3$) but enjoying tri-symmetry in the complex plane.
It leads me to a mechanical analog problem of a Newtonian particle on~$\C$ in a potential~$-V$.
I obtain the same action by plugging (\ref{ansatz7}) directly into~(\ref{fullaction}) with
$\diff\Sigma=*{\cal H}$ from~(\ref{ktorsion}).

For the case of $\frac{K}{H}=S^6=\frac{G_2}{\text{SU}(3)}$,
equation~(\ref{phieom}) produces in fact {\em all\/} $G$-equivariant Yang-Mills connections on 
$\R_\tau\times\sfrac{K}{H}$. On $\frac{\text{Sp}(2)}{\text{Sp}(1){\times}\text{U}(1)}$ and
$\sfrac{\text{SU}(3)}{\text{U}(1){\times}\text{U}(1)}$, however, the most general $G$-equivariant
connections involves two respective three complex functions of~$\tau$. The corresponding Newtonian
dynamics on $\C^2$ respective $\C^3$ is of similar type but constrained by the conservation
of Noether charges related to relative phase rotations of the complex functions.

\section{Seven dimensions: solutions}

Finite-action solutions require Newtonian trajectories between zero-potential critical 
points~$\hP$. With two exotic exceptions, $\diff V(\hP)=0=V(\hP)$ \ yields precisely the 
BPS configurations on~$\frac{K}{H}$:
\begin{itemize}
\addtolength{\itemsep}{-6pt}
\item $\hP=\eu^{2\pi\ic k/3}$ \quad with \ $V(\hP)=0$ \ \qquad for all values of~$\k$ and $k=0,1,2$
\item $\hP=0$ \qquad\quad with \ $V(\hP)=3{-}\k$ \quad vanishing only at $\k{=}3$
\end{itemize}
Kink solutions will interpolate between two different critical points, while bounces will return
to the critical starting point.
Thus for generic $\k$ values one may have kinks of `transversal' type, connecting two third roots of unity, 
as well as bounces. For $\k{=}3$ `radial' kinks, reaching such a root from the origin, may occur as well.
Numerical analysis reveals the domains of existence in~$\k$:
\begin{center}
\begin{tabular}{|l|ccccc|}
\hline
$\k$ interval & $(-\infty,-3]$ & $(-3,+3)$ & $+3$ & $(+3,+5)$ & $[+5,+\infty)$ \\
types of   & radial & transversal & radial & radial & -- \\[-4pt]
trajectory & bounce & kink        & kink   & bounce & -- \\
\hline
\end{tabular}
\end{center}

On the following page I display contour plots of the potential and finite-action trajectories for eight
choices of~$\k$. They reveal three special values of~$\k$:
At $\k{=}{-}3$ rotational symmetry emerges; this is a degenerate situation. At $\k{=}{-}1$ and at $\k{=}{+}3$,
the trajectories are straight, indicating integrability. Indeed, behind each of these two cases lurks 
a first-order flow equation, which originates from anti-self-duality and hence a particular $G_2$-structure~$\psi$.

Let me first discuss $\k{=}{+}3$.
For this value I find that
\be \label{gflow}
3\ddot\Phi=\sfrac{\pa V}{\pa\bar\Phi} \qquad\Longleftarrow\qquad
\sqrt{2}\dot\Phi=\pm\sfrac{\pa W}{\pa\bar\Phi} \qquad\text{with}\qquad
W=\sfrac13(\Phi^3{+}\bar\Phi^3)-|\Phi|^2\ ,
\ee
which is a gradient flow with a real superpotential~$W$, as
\be
V\=6\,\bigl|\sfrac{\pa W}{\pa\bar\Phi}\bigr|^2
\qquad\text{for}\quad \k=+3\ .
\ee
It admits the obvious analytic radial kink solution,
\be
\Phi(\tau)\=\eu^{\frac{2\pi\ic k}{3}}
\bigl(\sfrac12\pm\sfrac12\tanh\sfrac{\tau}{2\sqrt{3}}\bigr)\ .
\ee

What is the interpretation of this gradient flow in terms of the original
Yang-Mills theory? 
Demanding that the torsion in~(\ref{ktorsion}) comes from a $G_2$-structure,
$*{\cal H}=\diff\psi$, I am led to
\be
\psi \= \sfrac13\k\,\omega\wedge\diff\tau\ +\ \alpha\,\text{Im}\Omega
\qquad\Longrightarrow\qquad
\diff\psi\ \sim\ \k\,\text{Im}\Omega\wedge\diff\tau\ \sim\ \psi\wedge\diff\tau
\ee
where $\alpha$ is undetermined. This is a conformally parallel $G_2$-structure,
and (\ref{G2ASD}) quantizes the coefficients to $\alpha{=}1$ and $\k{=}3$, fixing
\be \label{str1}
\psi \= \omega\wedge\diff\tau\ +\ \text{Im}\Omega \=
r^{-3}\bigl(r^2 \omega\wedge\diff r+ r^3 \text{Im}\Omega \bigr) \= r^{-3}\psi_{\text{cone}}
\qquad\text{with}\quad \eu^\tau=r\ ,
\ee
where I displayed the conformal relation to the parallel $G_2$-structure on the cone over~$\sfrac{K}{H}$.

Alternatively, with this $G_2$-structure the 7 anti-self-duality equations~(\ref{G2ASD}) turn into
\be \label{flow}
\omega\lrcorner F\ \sim\ J_{ab}\,F_{ab}\=0 \und 
\dot{A}\ \sim\ \diff\omega\lrcorner F\ \sim\ e^a\,f_{abc}\,F_{bc}\ .
\ee
With the ansatz (\ref{ansatz7}), the first relation is automatic, and the second one indeed 
reduces to~(\ref{gflow}).
As a consistency check, one may verify that 
\be
\int_{\frac{K}{H}}\!\tr\,\{\om\we F\we F\}\ \propto\ W(\Phi)+\sfrac13\ .
\ee

\newpage

\begin{center}
\begin{minipage}{6cm}
\includegraphics[width=6cm]{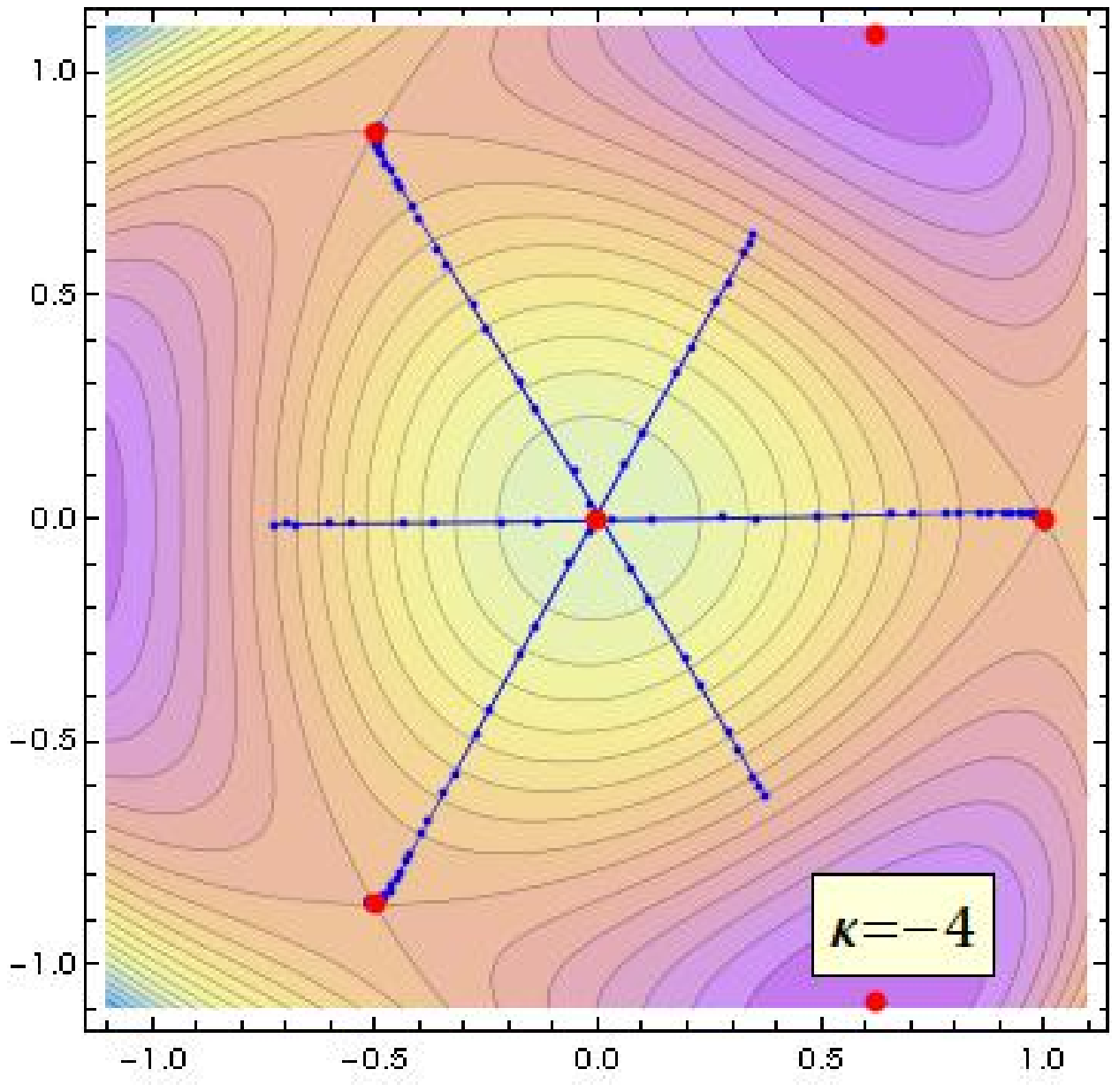}
\end{minipage}
\qquad
\begin{minipage}{6cm}
\includegraphics[width=6cm]{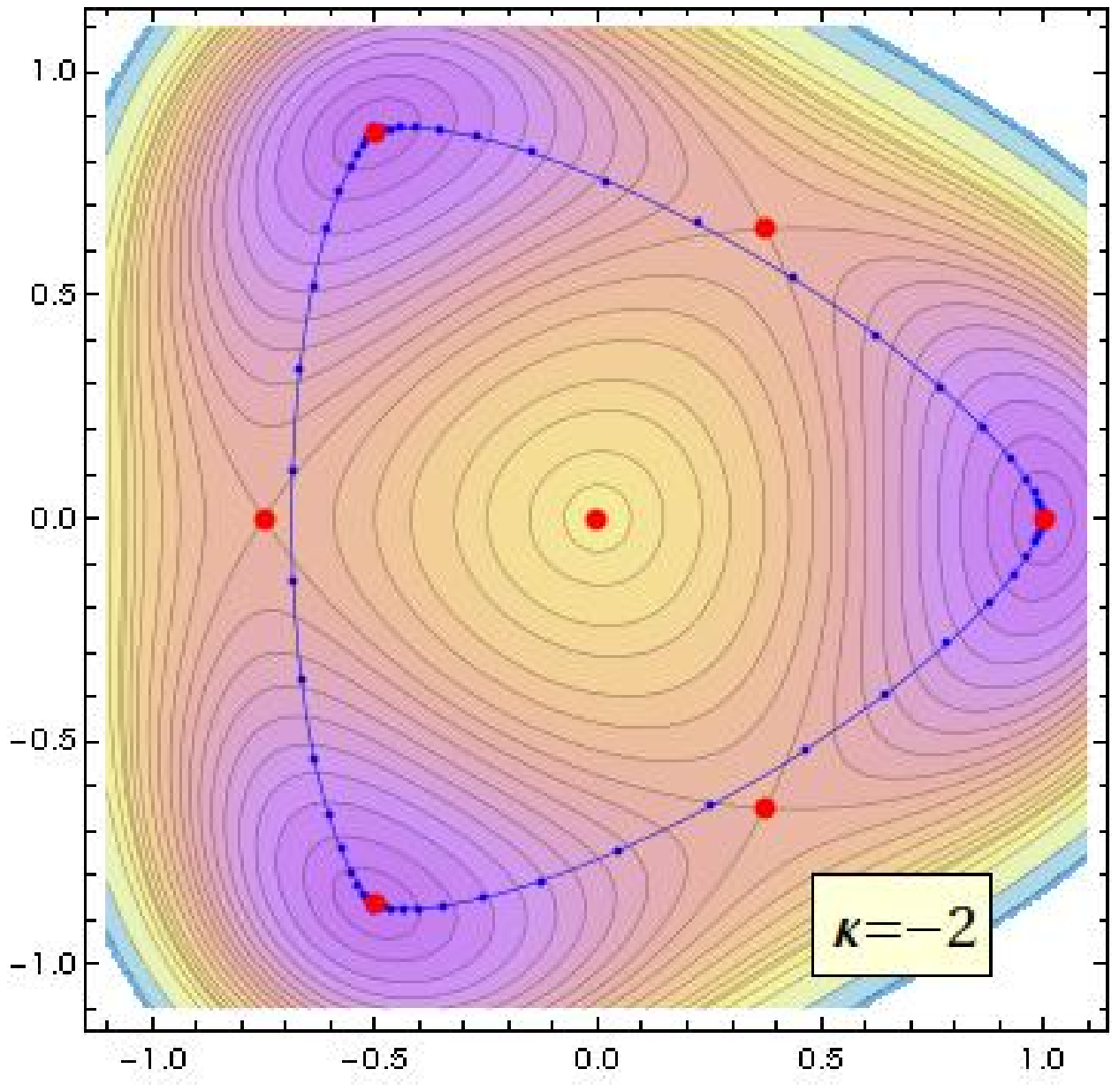}
\end{minipage}
\\[4pt]
\begin{minipage}{6cm}
\includegraphics[width=6cm]{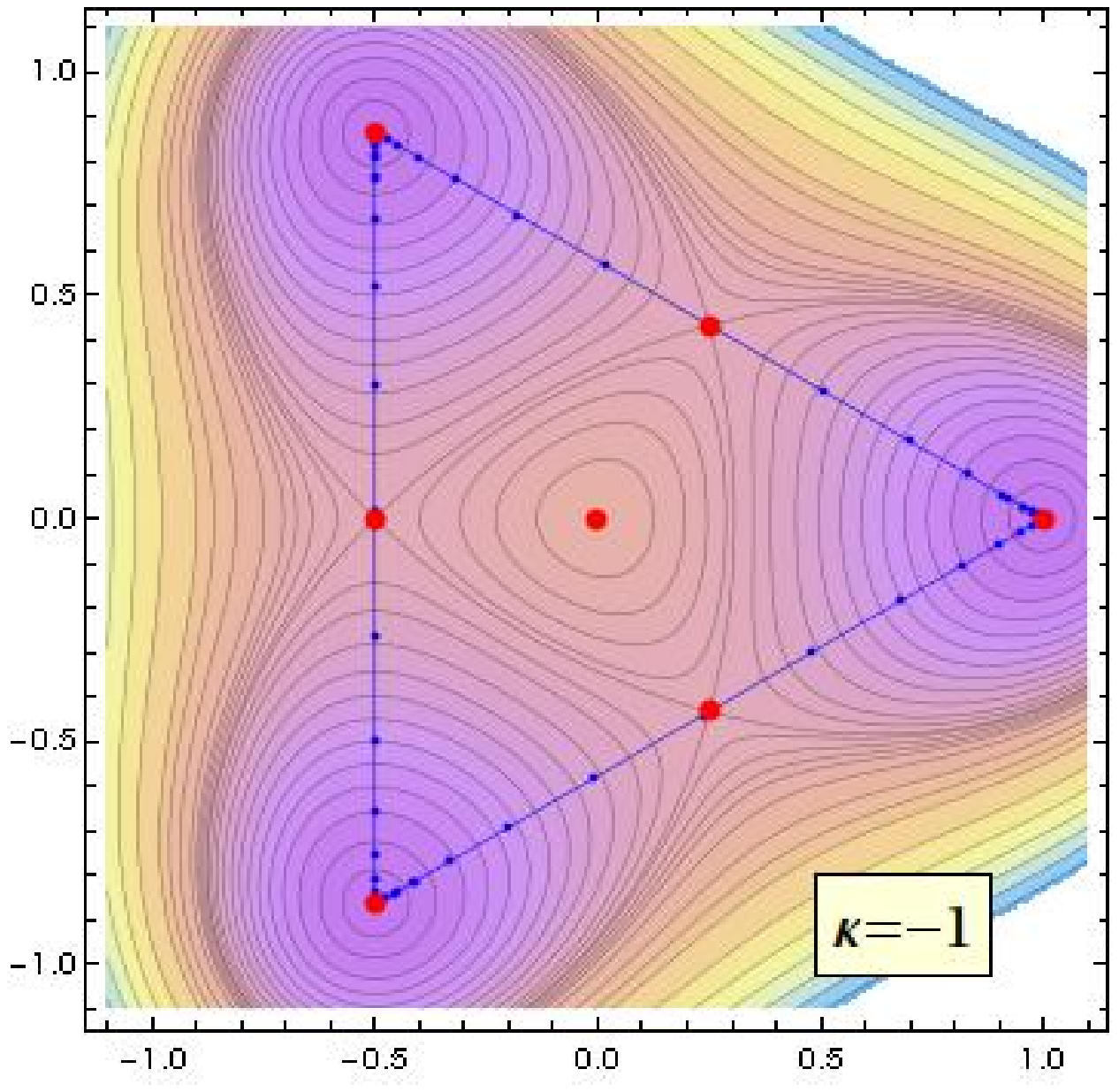}
\end{minipage}
\qquad
\begin{minipage}{6cm}
\includegraphics[width=6cm]{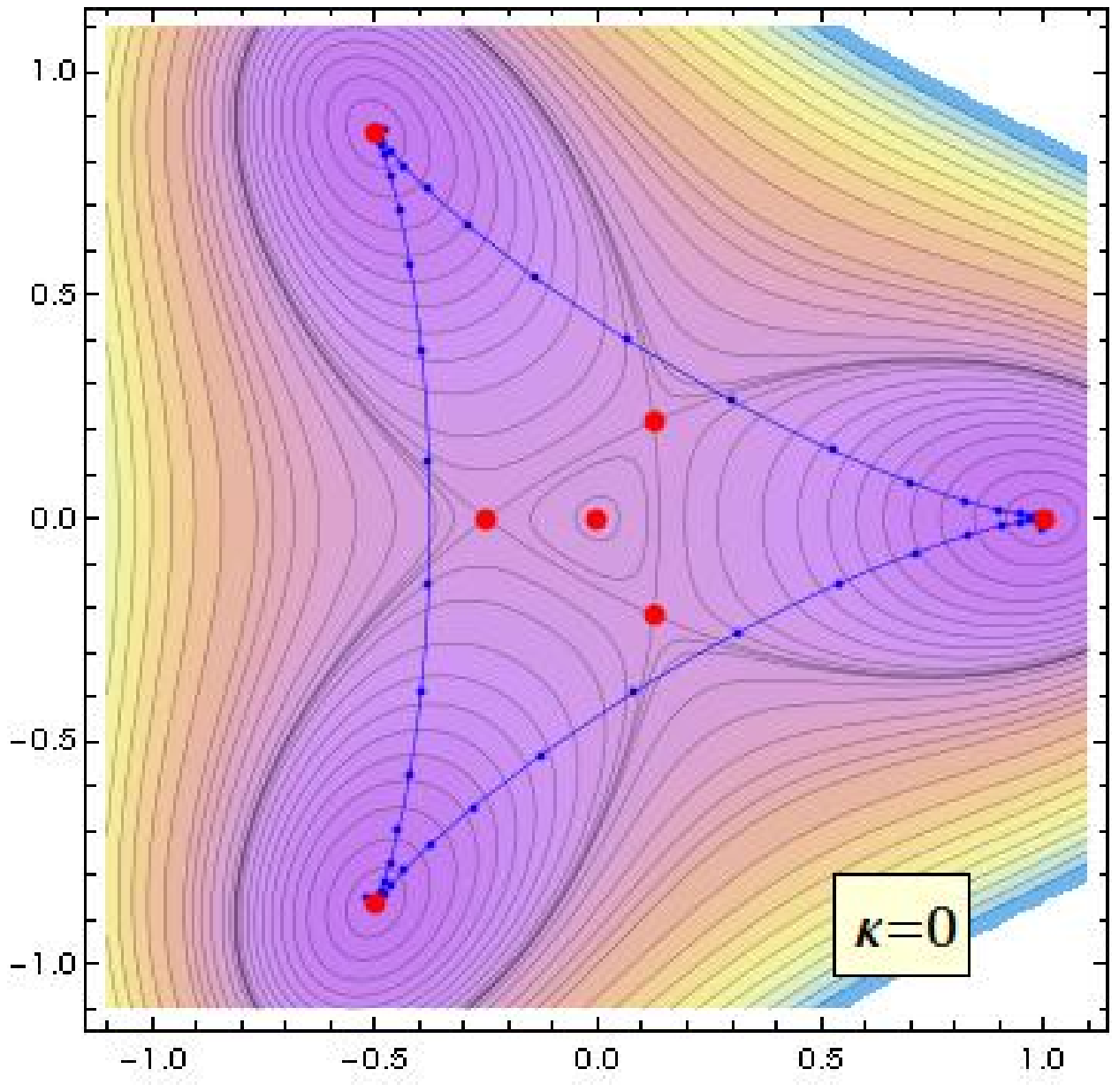}
\end{minipage}
\\[4pt]
\begin{minipage}{6cm}
\includegraphics[width=6cm]{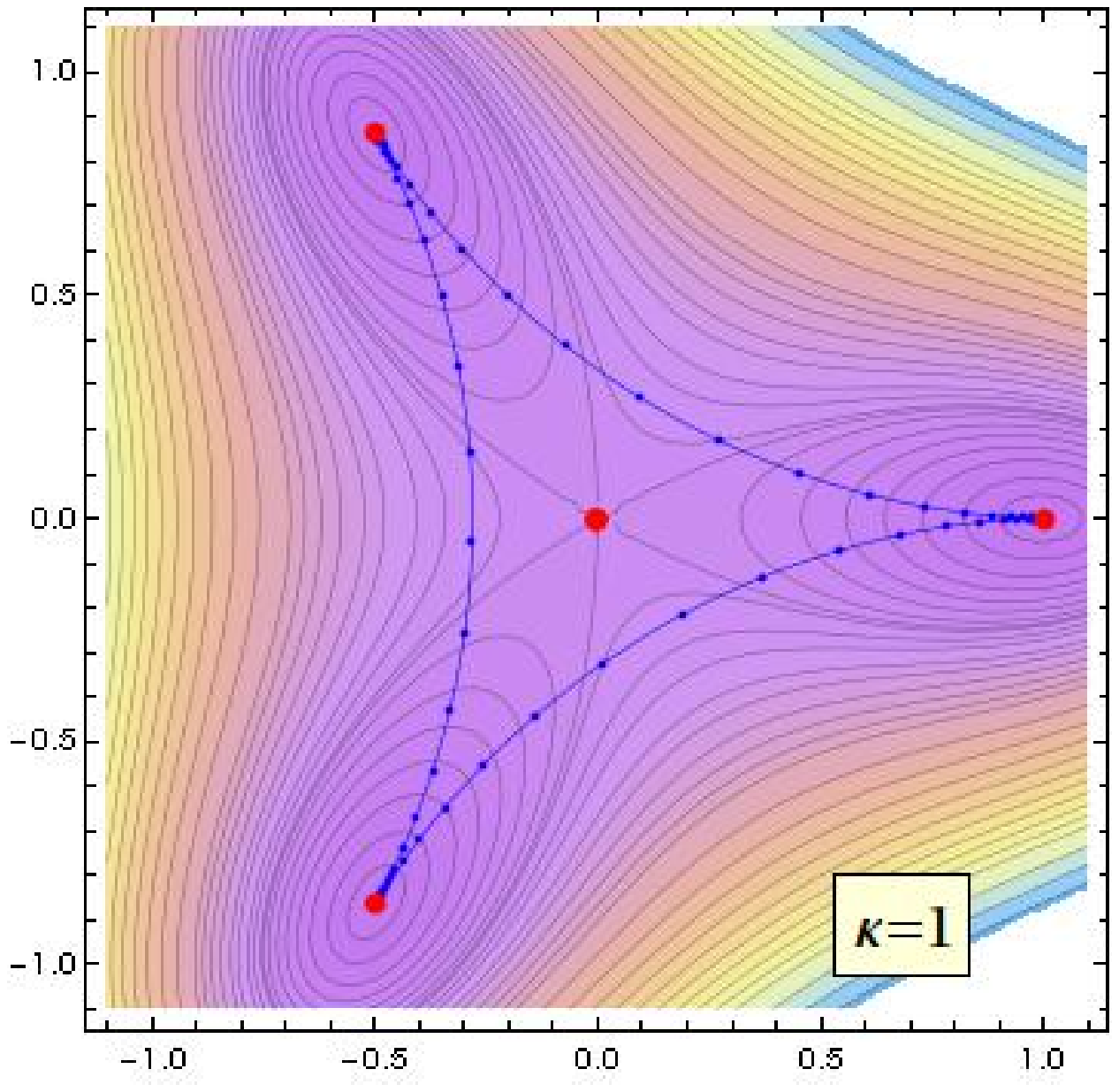}
\end{minipage}
\qquad
\begin{minipage}{6cm}
\includegraphics[width=6cm]{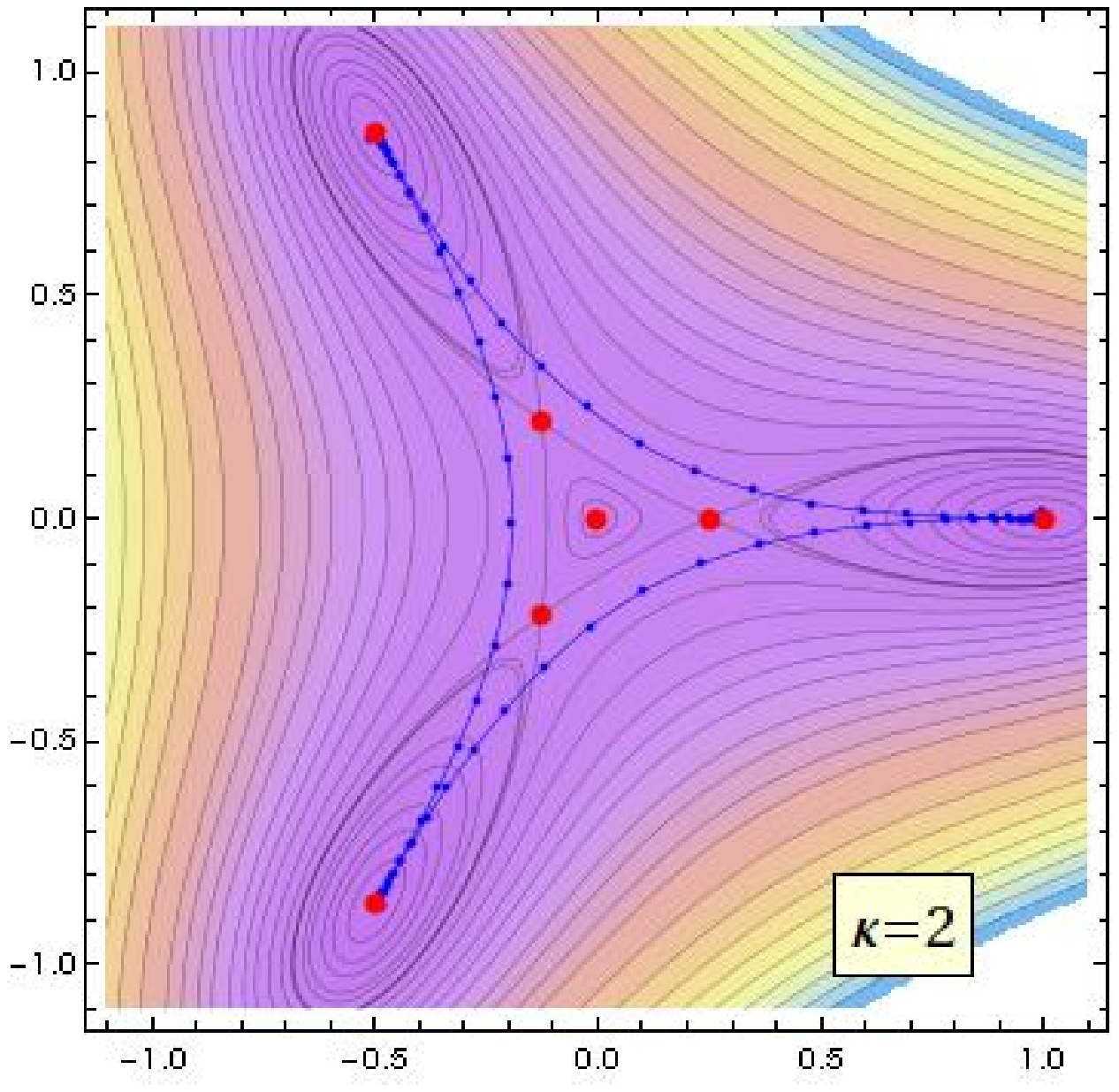}
\end{minipage}
\\[4pt]
\begin{minipage}{6cm}
\includegraphics[width=6cm]{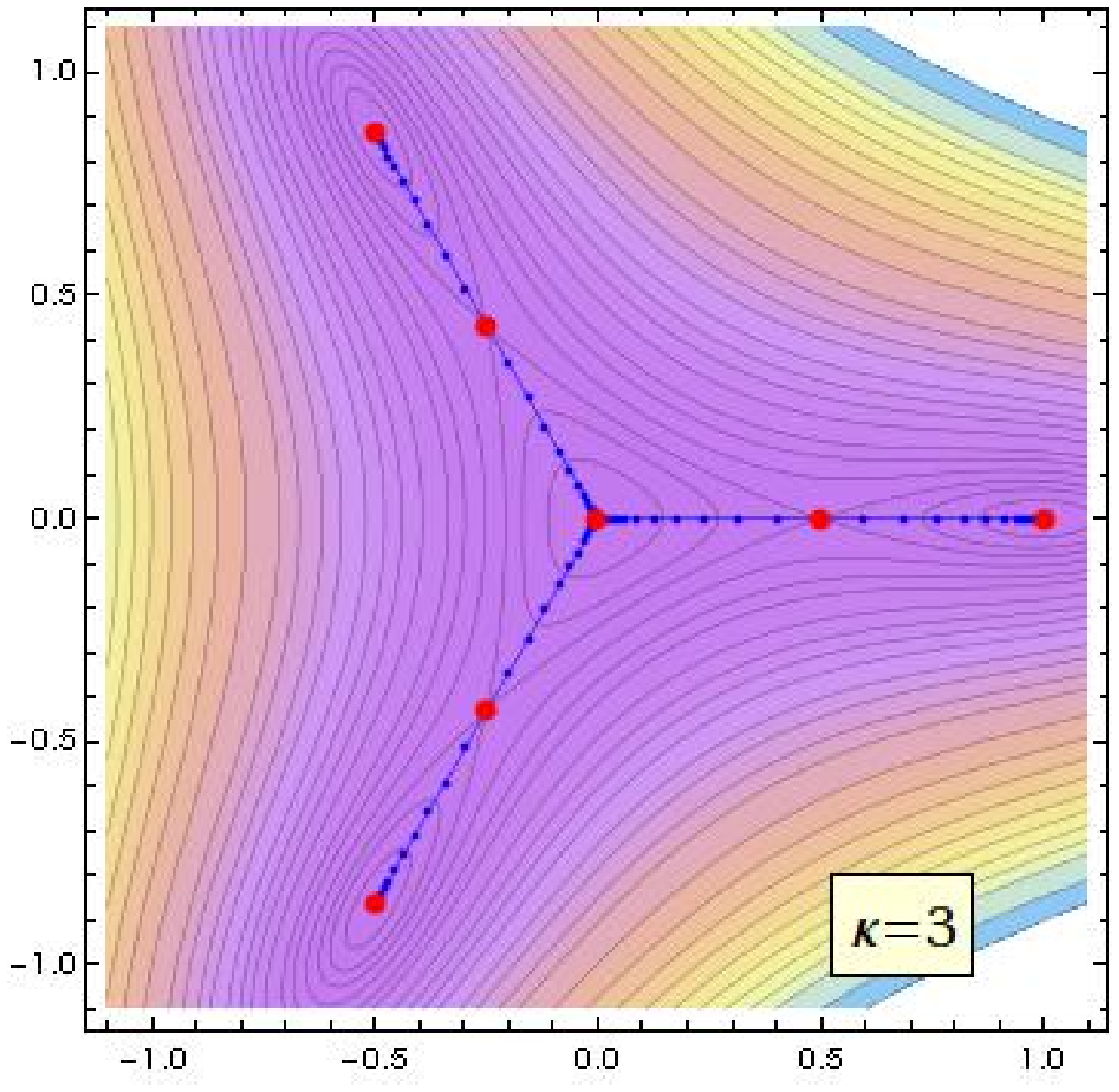}
\end{minipage}
\qquad
\begin{minipage}{6cm}
\includegraphics[width=6cm]{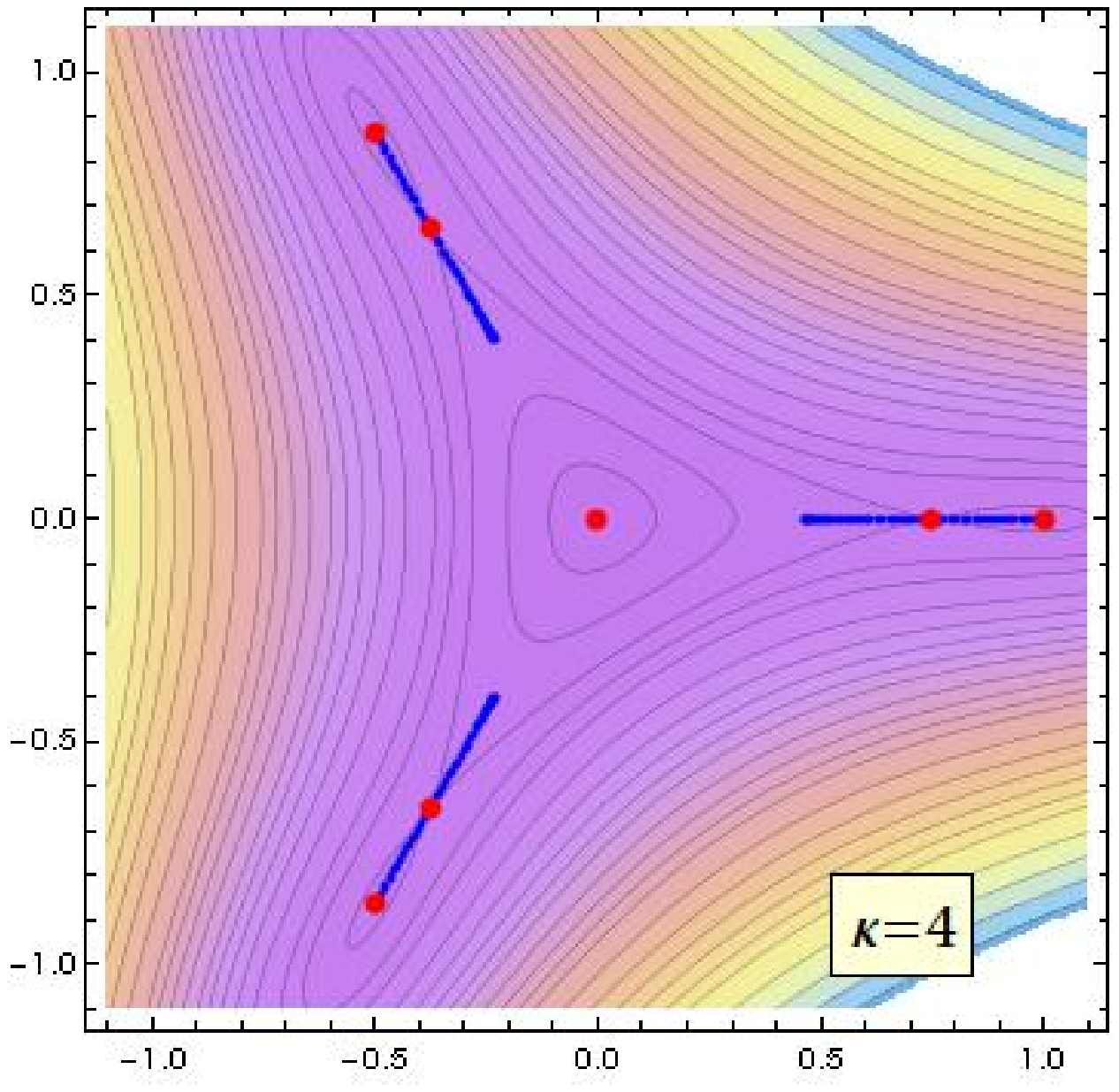}
\end{minipage}
\end{center}

\newpage

I now come to the other instance of straight trajectories, $\k{=}{-}1$.
For this value I find that
\be \label{hflow}
3\ddot\Phi=\sfrac{\pa V}{\pa\bar\Phi} \qquad\Longleftarrow\qquad
\sqrt{2}\dot\Phi=\pm\ic\,\sfrac{\pa H}{\pa\bar\Phi} \qquad\text{with}\qquad
H=\sfrac13(\Phi^3{+}\bar\Phi^3)-|\Phi|^2\ ,
\ee
which is a hamiltonian flow (note the imaginary multiplier!), 
running along the level curves of the function~$H$,
that is identical to~$W$.
It has the obvious analytic transverse kink solution,
\be
\Phi(\tau)\=-\sfrac12\pm\sfrac{\sqrt{3}}{2}\,\ic\,(\tanh\sfrac{\tau}{2})
\ee
and its images under the tri-symmetry action.

Have I discovered another hidden $G_2$-structure here?
Let me try the other obvious choice,
\be
\widetilde\psi \= \sfrac13\tilde\k\,\omega\wedge\diff\tau\ +\ \tilde\alpha\,\text{Re}\Omega
\qquad\Longrightarrow\qquad \diff\widetilde\psi\ \sim\ 
\widetilde\k\,\text{Im}\Omega\wedge\diff\tau\ +\ 2\tilde\alpha\,\omega\wedge\omega\ ,
\ee
with coefficients $\tilde\k$ and $\tilde\alpha$ to be determined.
It has not appeared in my table in Section~2, but obeys $\diff*\widetilde\psi=0$, which is known
as a {\em cocalibrated\/} $G_2$-structure. But can it produce the proper torsion,
\be
\diff\widetilde\psi\wedge F\ \sim\ 
\bigl(\widetilde\k\,\text{Im}\Omega\wedge\diff\tau+2\tilde\alpha\,\omega\wedge\omega\bigr)\wedge F
\ \buildrel{!}\over{=}\ -\text{Im}\Omega\wedge\diff\tau\wedge F\ \ ?
\ee
Employing the anti-self-duality with respect to~$\tilde\psi$,
\be
*\widetilde\psi\wedge F \=0 \qquad\Longrightarrow\qquad 
\omega\wedge\omega\wedge F \= 2\,\text{Im}\Omega\wedge\diff\tau\wedge F\ ,
\ee
it works out, adjusting the coefficients to $\tilde\k{=}3$ and $\tilde\alpha{=}{-}1$.
Hence, my cocalibrated $G_2$-structure
\be \label{str2}
\widetilde\psi\= \omega\wedge\diff\tau\ -\ \text{Re}\Omega
\ee
is responsible for the hamiltonian flow.

To see this directly, I import~(\ref{str2}) into~(\ref{G2ASD}) and get
\be
J_{ab}\,F_{ab}\=0 \und \dot{A}_a\ \sim\ [J\,f]_{abc}\,F_{bc}\ .
\ee
Again, the ansatz (\ref{ansatz7}) fulfills the first relation, but the second one nicely
turns into~(\ref{hflow}). \\[8pt]

\begin{minipage}{10cm}
\includegraphics[width=10cm]{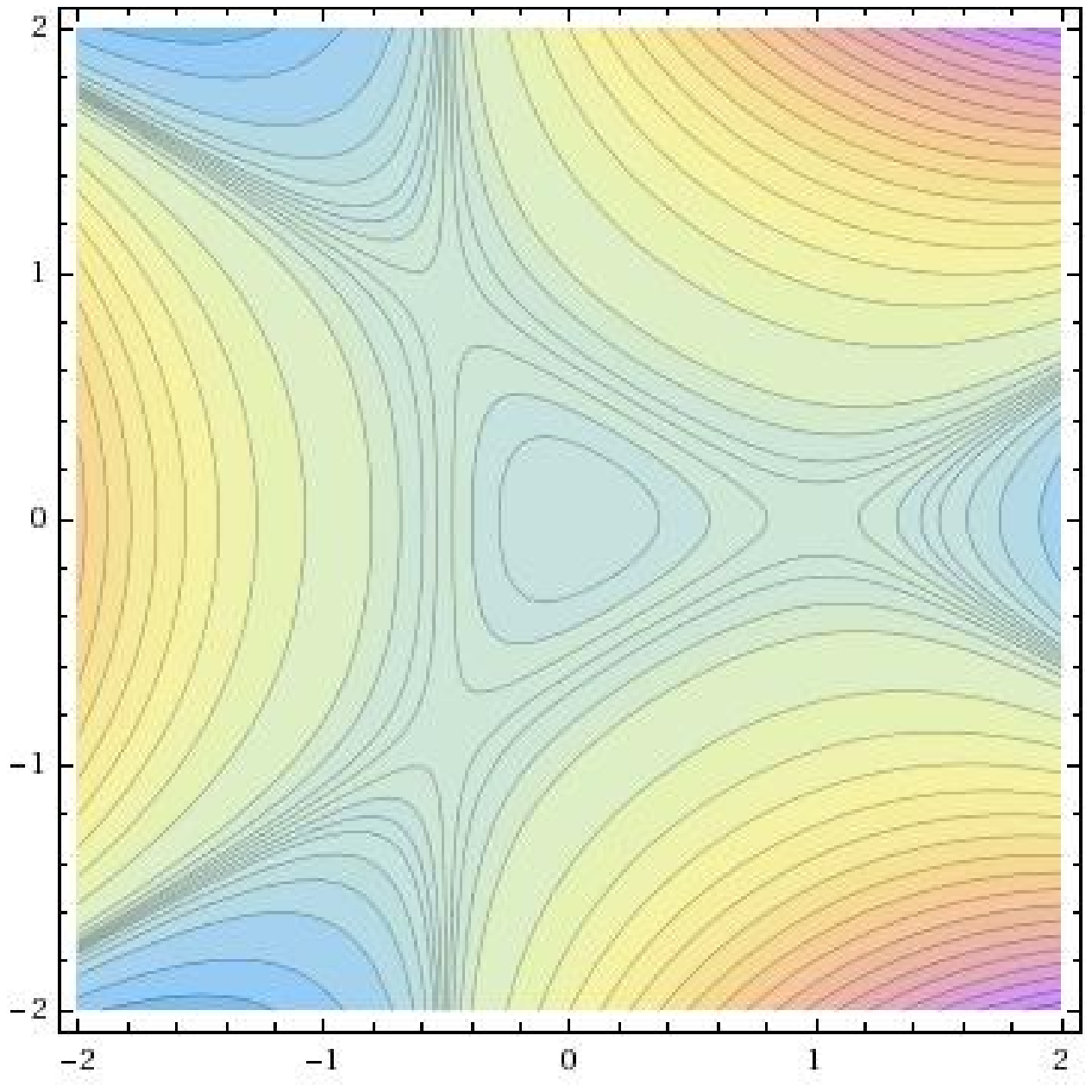} \\
\ph\quad contours of superpotential/hamiltonian $W(\Phi)\equiv H(\Phi)$ 
\end{minipage}
\qquad
\begin{minipage}{5cm}
\includegraphics[width=5cm]{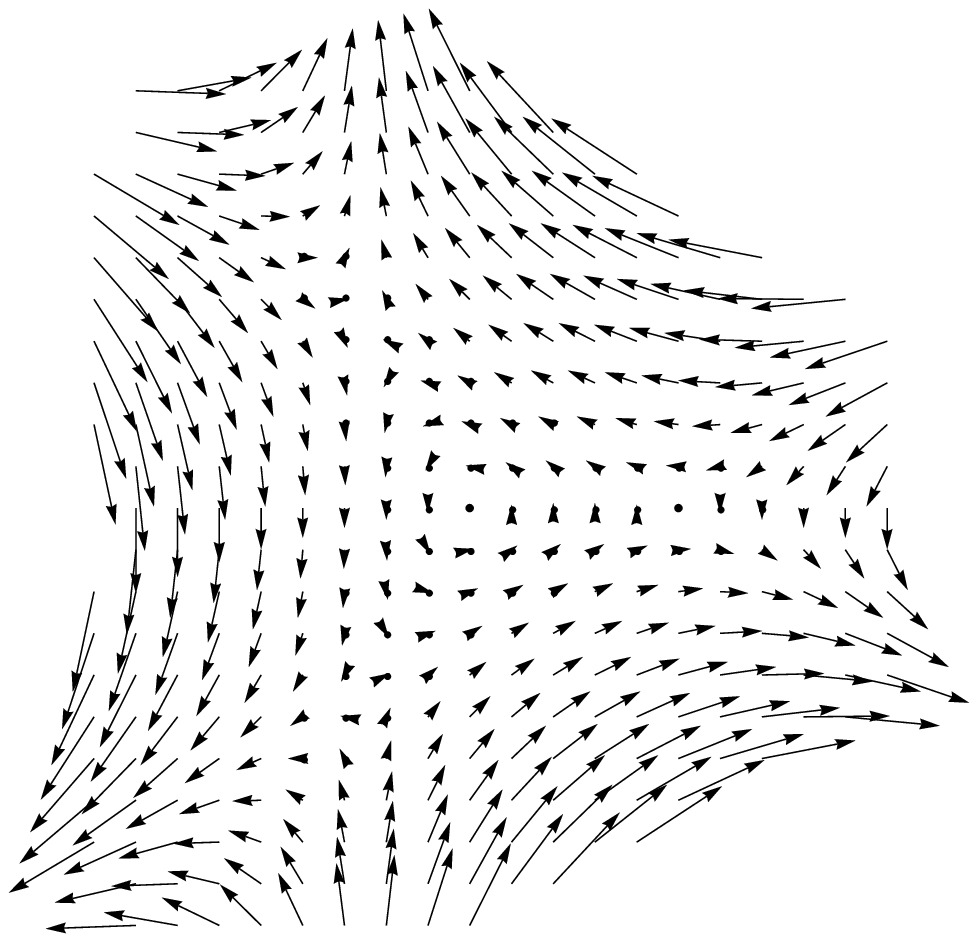} \\[-3mm]
\ph\quad hamiltonian vector field\\
\includegraphics[width=4.5cm]{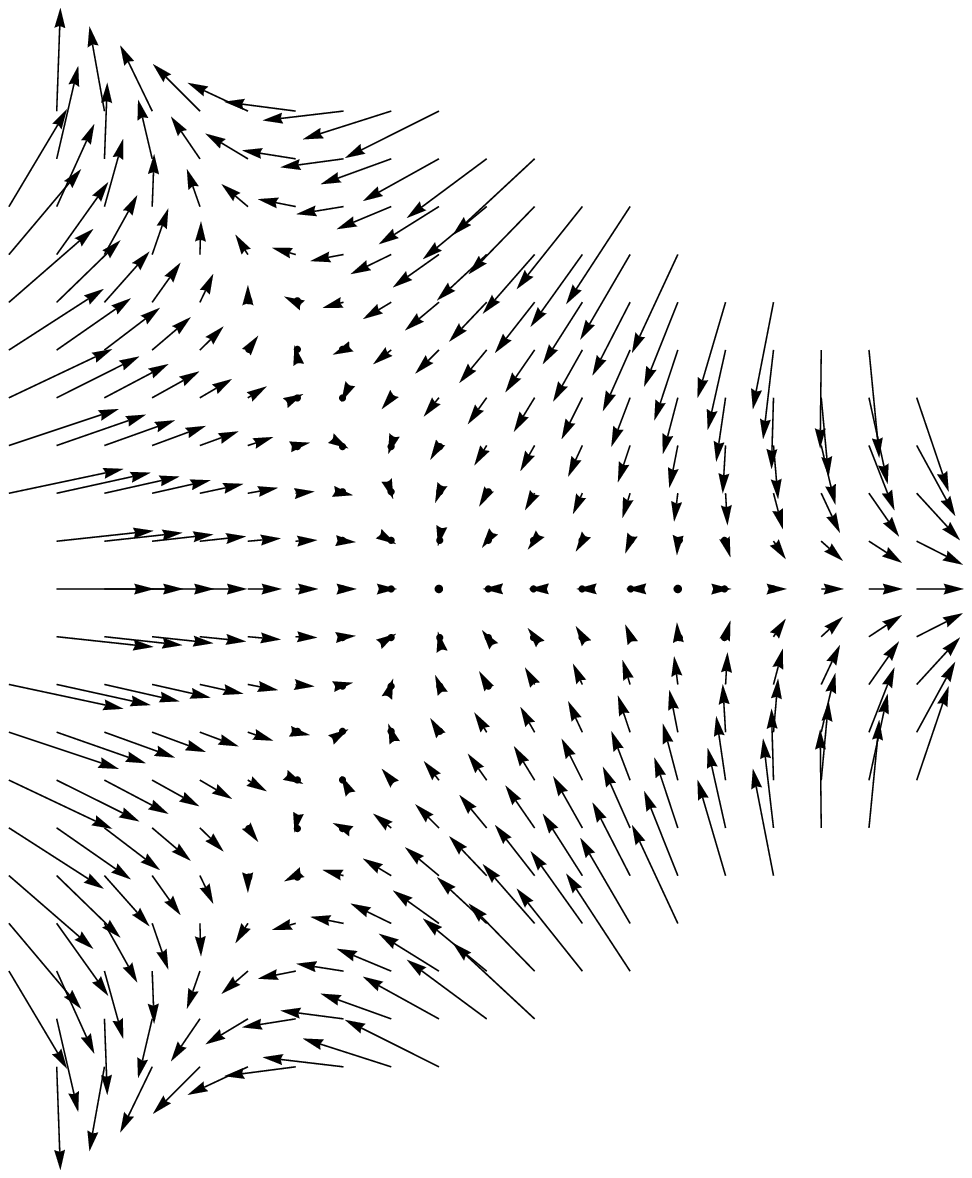} \\[-5mm]
\ph\quad gradient vector field
\end{minipage}

The story has an eight-dimensional twist, which can be inferred from the diagram
in Section~2. There it is indicated that my cylinder is embedded into an 8-manifold~$M^8$ 
equipped with a parallel Spin(7)-structure~$\Sigma$. It can be regarded as the cylinder 
over the cone over $\sfrac{K}{H}$. The four-form~$\Sigma$ descends to the cocalibrated
$G_2$-structure~$\widetilde\psi$, while $\psi$ is obtained by reducing to the cone and
applying a conformal transformation.

The anti-self-duality condition on~$M^8$ represents 7 relations, which project $F_8$ to the 
21-dimensional $-1$ eigenspace of $*(\Sigma\wedge\cdot)$. 
Contrary to the $G_2$ situation~(\ref{flow}), where 7 anti-self-duality equations split to 
6 flow equations and the supplementary condition $\omega\lrcorner F{=}\,0$, for Spin(7) the count 
precisely matches, as I have also 7 flow equations. Indeed, there is equivalence:
\be
*_8 F_8\=-\S\wedge F_8
\qquad\Longleftrightarrow\qquad
\sfrac{\pa A_7(\sigma)}{\pa\sigma}\=*_7\bigl(\diff\psi\wedge F_7(\sigma)\bigr)\ .
\ee

\section{Partial summary}

\begin{minipage}{3cm}
\begin{flushright}
\includegraphics[width=2cm]{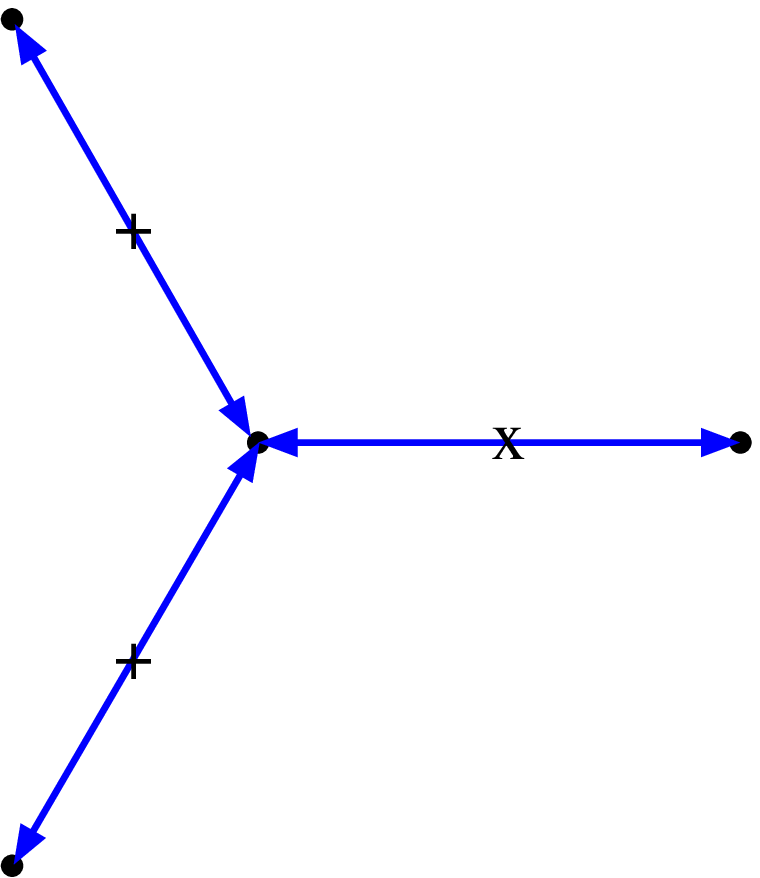}
\end{flushright}
\end{minipage}
\hfill
\begin{minipage}{3cm}
\begin{flushleft}
\includegraphics[width=2cm]{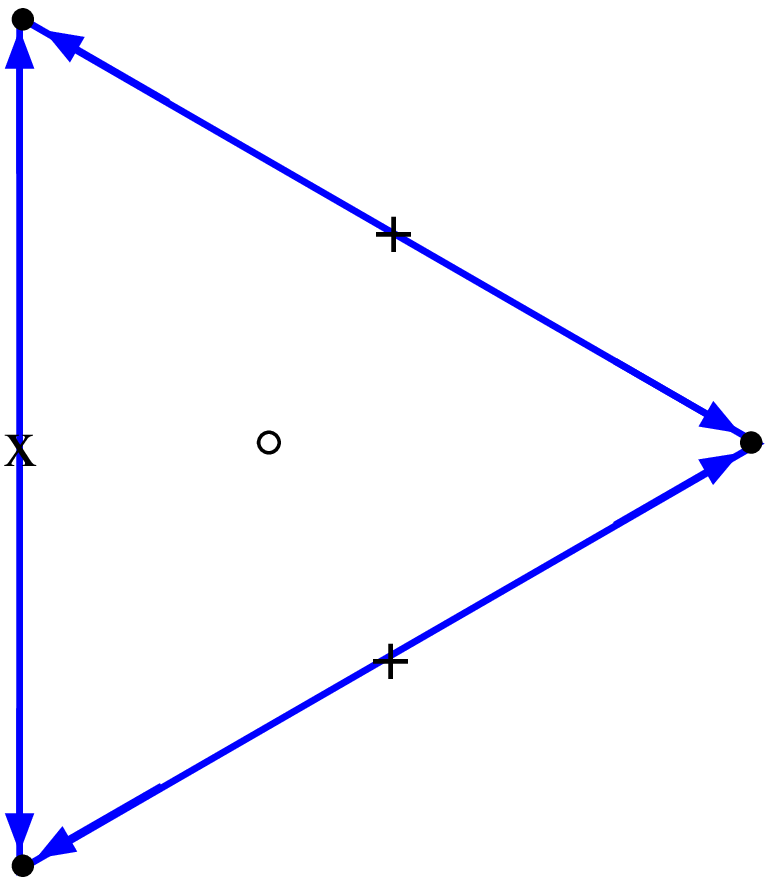}
\end{flushleft}
\end{minipage}
\vskip-2.5cm
\hfil Let me schematically sum up the construction.\hfil\\[2pt]
\begin{eqnarray*}
\S\we F_8\!\!\!&=&\!\!\! -*_8 F_8 \\
\swarrow\qquad\qquad&&\qquad\qquad\searrow \\
\psi\we F\=-*_7 F \qquad\qquad\quad
\text{on}\ \ \R\!\!\!\!\!&\times&\!\!\!\!\!\sfrac{K}{H}
\qquad\qquad\qquad\ \widetilde\psi\,{\we}\,F\=-*_7 F \\
\dot{A}_a\ \sim\ f_{abc}\,F_{bc}
\qquad\qquad\qquad&&\qquad\qquad\quad
\dot{A}_a\ \sim\ [\J f]_{abc}\,F_{bc}\\[4pt]
\big\downarrow\quad\qquad
\text{ansatz}\quad
A=\!\!\!\!& e^i&\!\!\!\!\!\! I_i+e^a[\Phi\,I]_a
\qquad\quad\ \big\downarrow\\[4pt]
\sqrt{2}\dot\Phi=\pm\sfrac{\pa W}{\pa\bar\Phi}
\qquad\qquad\qquad&&\qquad\qquad\qquad
\sqrt{2}\dot\Phi=\pm{\ic}\,\sfrac{\pa H}{\pa\bar\Phi}\\[4pt]
\Big\downarrow\qquad
W \= \sfrac13 (\Phi^3{+}\!\!\!&\bar\Phi^3&\!\!\!)-|\Phi|^2 \= H
\qquad\Big\downarrow
\end{eqnarray*}
$\ph\qquad\qquad\quad 
F(\tau)\= \diff\tau{\we}e^a\,[\dot\Phi\,I]_a\ +\
\sfrac12 e^a{\we}e^b\bigl\{
(|\Phi|^2{-}1)\,f^i_{ab}\,I_i\ +\ [(\bar\Phi^2{-}\Phi)\,f]_{abc}\,I_c\bigr\}$
\\[12pt]
are $G_2$-instantons for Yang-Mills with torsion \qquad
$D*\!F+(*{\cal H})\we F\=0$ \\[8pt]
from \qquad
$S[A]\=\int_{\R\times\frac{K}{H}}\!\tr\bigl\{
F\we*\!F\ +\ \sfrac13\k\,\om\we\diff\tau\we F\we F\bigr\}
\qquad\text{with}\ \ \k=+3 \ \text{or} \ {-}1$ \\[4pt]
and obey gradient/hamiltonian flow equations for \ \
$\int_{\frac{K}{H}}\!\tr\,\{\om\we F\we F\}\ \propto\
W(\Phi)+\sfrac13$\ .

\end{document}